%% file: miccai2022.tex
\documentclass[runningheads]{llncs}

\usepackage[T1]{fontenc}
\usepackage{graphicx}
\usepackage{microtype}
\usepackage{subfigure}
\usepackage{booktabs} 
\usepackage{url}
\usepackage{amsmath}
\usepackage{amssymb}
\usepackage{latexsym}
\usepackage{array}
\usepackage{adjustbox}
\usepackage{multirow}
\usepackage{hyperref}
\usepackage{longtable}
\usepackage{soul}
\usepackage[inline]{enumitem}
\usepackage{xspace}
\usepackage{wrapfig}
\usepackage{makecell}
\input{content/tables}
\begin{document}
\title{Scale-Equivariant Unrolled Neural Networks for Data-Efficient Accelerated MRI Reconstruction}
\titlerunning{Scale-Equivariant Unrolled Neural Networks}
%
\author{Beliz Gunel\inst{1,2} \and
Arda Sahiner\inst{1,2} \and
Arjun D. Desai \inst{1} \and
Akshay S. Chaudhari \inst{1} \and
Shreyas Vasanawala \inst{1} \and
Mert Pilanci \inst{1} \and
John Pauly \inst{1}}
\authorrunning{Gunel \& Sahiner et al.}
\institute{Stanford University, CA, 94305, USA \and Equal Contribution
\email{\{bgunel,sahiner,arjundd,akshaysc,vasanawala,akshaysc,pilanci,pauly\}@stanford.edu}}

%

%
\maketitle              
\begin{abstract}
Unrolled neural networks have enabled state-of-the-art reconstruction performance and fast inference times for the accelerated magnetic resonance imaging (MRI) reconstruction task. However, these approaches depend on fully-sampled scans as ground truth data which is either costly or not possible to acquire in many clinical medical imaging applications; hence, reducing dependence on data is desirable. In this work, we propose modeling the proximal operators of unrolled neural networks with \textit{scale-equivariant} convolutional neural networks in order to improve the data-efficiency and robustness to drifts in scale of the images that might stem from the variability of patient anatomies or change in field-of-view across different MRI scanners. Our approach demonstrates strong improvements over the state-of-the-art unrolled neural networks under the same memory constraints both with and without data augmentations on both in-distribution and out-of-distribution scaled images without significantly increasing the train or inference time.


\end{abstract}

\section{Introduction}
\label{section:introduction}
\input{content/1_introduction.tex}

\section{Related Work}
\label{section:related_work}
\input{content/2_related_work.tex}

\section{Background and Preliminaries}
\label{section:background}
\input{content/3_background_and_preliminaries.tex}

\section{Methods}
\label{section:methods}
\input{content/4_methods.tex}

\section{Experiments and Results}
\label{section:experiments}
\input{content/5_experiments.tex}

\section{Conclusion}
\label{section:conclusion}
\input{content/6_conclusion.tex}

\section{Acknowledgements}
\label{section:acknowledgements}
\input{content/7_acknowledgements.tex}
\bibliography{miccai2022}
\bibliographystyle{splncs04}





\end{document}

%% file: content/tables.tex
\newcommand{\insertscalevsvanilla}{
    \begin{table}[t!]
    \caption{Scale-Equivariant unrolled networks (Scale-Eq) outperform the state-of-the-art unrolled networks (Vanilla) both on in-distribution (None) and scaled images with scaling factor $a=0.9$ (Scaled), both for zero-initialized and tuned step size initializations $\eta_0$ in the data-limited regime. Both Scale-Eq and Vanilla networks use appropriate scale data augmentations during training, as noted by +. Scale-Eq+ for tuned $\eta_0$ uses $B=2$.}
    \label{table:maintable}
        \begin{center}
\begin{tabular}{cccccc}
\toprule
&&\multicolumn{2}{c}{Original} & \multicolumn{2}{c}{Scaled (a=0.9)}\\
Model & $\eta_0$ & SSIM & cPSNR (dB) &  SSIM &  cPSNR (dB) \\
\midrule
Vanilla+ & zero-init & 0.864 (0.012) & 35.77 (0.50) & 0.919 (0.013) & 39.03 (0.46)\\
Scale-Eq+ &  & \textbf{0.900 (0.003)} & \textbf{36.44 (0.49)}  & \textbf{0.944 (0.006)} & \textbf{40.45 (0.57)} \\
\cmidrule(lr){1-6}
Vanilla+ & tuned & 0.902 (0.008) & 35.59 (0.40)  & 0.952 (0.004)  & 40.15 (0.41) \\
Scale-Eq+ &  & \textbf{0.911 (0.008)} & \textbf{35.83 (0.46)}  & \textbf{0.956 (0.001)} & \textbf{40.97 (0.65)}\\
\bottomrule
\end{tabular}
        \end{center}
    \end{table}
}

\newcommand{\insertstepsize}{
    \begin{table}[t!]
    \caption{Scale-Equivariant unrolled neural networks (Scale-Eq) is less sensitive to the learnable step size initializations $\eta_0$ than the state-of-the-art unrolled neural networks (Vanilla), which is desirable since this parameter is highly dataset specific and needs to be tuned carefully. Scale-Eq consistently outperforms Vanilla across all initializations.}
    \label{table:stepsize}
        \begin{center}
\begin{tabular}{c|cc|cc||cc|cc}
\toprule
& \multicolumn{2}{c}{SSIM} & \multicolumn{2}{c}{cPSNR (dB)} & \multicolumn{2}{c}{SSIM ($a$=0.9)} & \multicolumn{2}{c}{cPSNR (dB) ($a$=0.9)} \\
$\eta_0$ & Vanilla & Scale-Eq &  Vanilla & Scale-Eq & Vanilla & Scale-Eq & Vanilla & Scale-Eq  \\
\midrule
2.5  & 0.872 & 0.880  & 36.84 & 37.06 & 0.878 & 0.936 & 39.23 & 40.49 \\
2 &  0.883 &0.885 & 36.77 & 37.03 & 0.923 & 0.942 & 39.61 & 40.38 \\
1.5  &0.878&0.886& 36.97 & 37.02 & 0.907&0.938 & 39.77&40.34\\
1 & 0.876 & 0.877 & 37.00 & 37.11  & 0.925 & 0.932 & 40.08 & 40.59  \\
0.5 &0.877&0.870 & 37.00 &37.00 &0.895&0.928 & 39.49&40.20\\
0 &0.863&0.875 & 36.32&36.89 &0.912&0.926 & 38.76&39.92\\
\midrule
\midrule
Avg & 0.875 & \textbf{0.879}& 36.82 & \textbf{37.02} &  0.907 &  \textbf{0.934} & 39.49 &  \textbf{40.32} \\
Std & 0.006 & \textbf{0.005} &  0.24 & \textbf{0.07} & 0.016 & \textbf{0.006} & 0.42 & \textbf{0.22}\\
\bottomrule
\end{tabular}
        \end{center}
    \end{table}
}

\newcommand{\insertmraugmentconfig}{
    \begin{table}[t!]
    \caption{Data augmentation configuration for the experiments shown in Table 2. $p$ is the effective probability of applying an augmentation, which is equivalent to the base probability multiplied by the weighting factor in the MRAugment framework \cite{Fabian2021DataAF}.}
    \label{table:aug-hyperparams}
    \begin{center}
    \begin{tabular}{llcc}
        \toprule
        \textbf{Transform} & \textbf{Parameters} & $\mathbf{p}$ \\
        \midrule
         H-Flip & N/A & 0.275 \\
         V-Flip & N/A & 0.275 \\
         $k \times 90^\circ$ rotation & $k \in \{2\}$ & 0.275 \\
         Rotation & [-180$^\circ$, 180$^\circ$] & 0.275 \\
         Translation & [-10\%, 10\%] & 0.55 \\
         Scale & [0.75, 1.25] & 0.55 \\
         Shear & [-15$^\circ$, 15$^\circ$] & 0.55 \\
         \bottomrule
    \end{tabular}
    \end{center}
\end{table}
}

\newcommand{\insertallequivariance}{
    \begin{table}[t!]
    \caption{Comparison of Scale-Equivariant, Rotation-Equivariant, and Vanilla unrolled networks using the state-of-the-art data augmentation pipeline MRAugment \cite{Fabian2021DataAF} during training in the data-limited regime. Learned step sizes $\eta_0$ are zero-initialized for all networks; and Scale-Eq uses $B=2$ for the order of the kernel basis. Scale-Eq-MRAugment outperforms other models, while Rotation-Eq-MRAugment outperforms Vanilla-MRAugment in cPSNR and does comparably in SSIM.}
    \label{table:allequivariance}
        \begin{center}
\begin{tabular}{lccc}
\toprule
Model& SSIM &  cPSNR (dB)\\
\midrule
Vanilla-MRAugment  & 0.877 (0.016) & 34.27 (0.31)\\
Scale-Eq-MRAugment & \textbf{0.909 (0.007)} & \textbf{36.37 (0.47)}  \\
Rotation-Eq-MRAugment  &  0.873 (0.008) & 36.30 (0.50)  \\
\bottomrule
\end{tabular}
        \end{center}
    \end{table}
}

\newcommand{\insertbenchmark}{
    \begin{table}[t!]
    \caption{Memory (number of trainable parameters), training time, and inference time comparison of Scale-Equivariant and Vanilla unrolled neural networks with appropriate scale data augmentations on a Titan RTX GPU. Scale-Eq+ does not significantly increase the training or inference time over Vanilla+ under the same memory constraints.}
    \label{table:benchmark}
        \begin{center}
\begin{tabular}{l|c|c|c}
\toprule
Model & $\#$ Params & \makecell{Avg Training \\ Time/Iter (s)} & \makecell{Avg Inference \\ Time/Slice (s)} \\
\midrule
Vanilla+ & 434,415 & 0.18 & 0.046\\
Scale-Eq+ & 434,185 & 0.2172 & 0.058\\
\bottomrule
\end{tabular}
        \end{center}
    \end{table}
}

\newcommand{\insertequiverror}{
\begin{figure}[t!]
    \centering
    \includegraphics[width=0.6\textwidth]{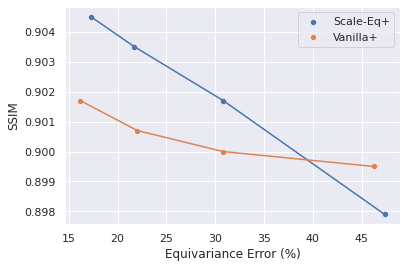}
    \caption{Sensitivity analysis comparing a vanilla network (Vanilla+) to a scale-equivariant network (Scale-Eq+) with varied equivariance error, where equivariance error is measured after training over the test set, and is controlled by manipulating the desired scale \(a\) to which to be equivariant. Network architectures which better enforce equivariance generally provide better performance, though the method of enforcement (augmentation v.s. equivariance + augmentation) also impacts performance.}
    \label{fig:equiv_error}
\end{figure}
}

\newcommand{\insertreconexamples}{
\begin{figure}[t!]
    \centering
    \includegraphics[width=0.6\textwidth]{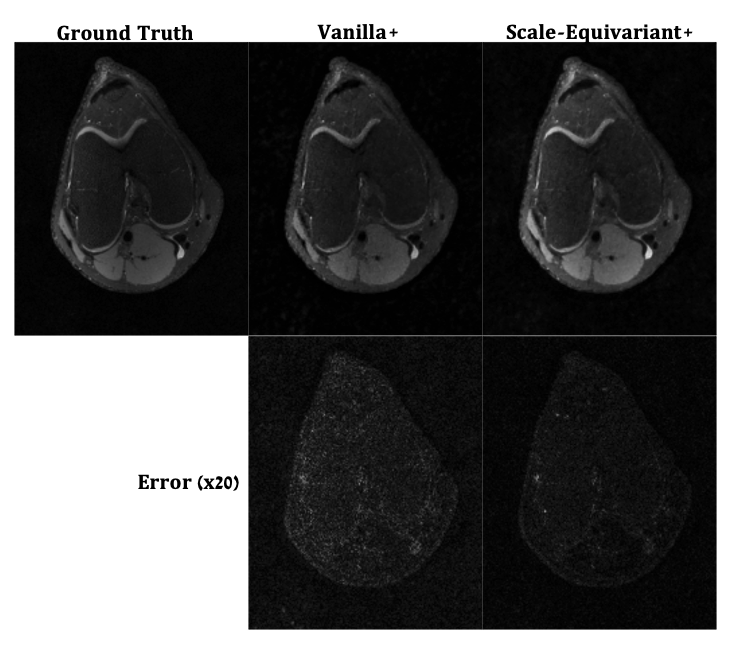}
    \caption{Reconstruction examples for Vanilla+ vs. Scale-Equivariant+ on in-distribution (Original) test scans. Error maps which show the difference between model reconstructions and the fully-sampled ground truth (x20) demonstrate that Scale-Equivariant+ leads to better reconstructions than Vanilla+.}
    \label{fig:recon_examples}
\end{figure}
}

%% file: content/1_introduction.tex
Magnetic resonance imaging (MRI) is a medical imaging modality that enables non-invasive anatomical visualization with great soft-tissue contrast and high diagnostic quality. Although the MRI data acquisition process can suffer from long scan durations, it can be accelerated by undersampling the requisite spatial frequency measurements, referred to as \textit{k-space} measurements. As the k-space measurements are subsampled below the Nyquist rate, reconstructing the underlying images without aliasing artifacts from these measurements is an ill-posed problem. To tackle this, previous approaches have leveraged prior knowledge of the true underlying solution in the form of regularization -- most notably enforcing sparsity in the Wavelet domain, referred to as \textit{compressed sensing} \cite{cs-mri}. However, these approaches suffer from long reconstruction times due to their iterative nature and can require parameter-specific tuning \cite{lustig2007sparse}.\\\\
Unrolled neural networks \cite{Sriram2020EndtoEndVN,sandino2020compressed,Aggarwal_Mani_Jacob_2019} have recently been shown to offer state-of-the-art reconstruction performance and faster reconstruction times compared to the traditional iterative methods, enabling higher acceleration factors for clinical applications \cite{Chaudhari2021}. However, these approaches still depend on fully-sampled scans as ground truth data which is either costly or not possible to acquire in many clinical medical imaging applications; hence, reducing dependence on data, referred to as \textit{data-efficiency}, is desirable. Towards the goal of improving data-efficiency, besides the prior proposals which leverage prospectively undersampled (unsupervised) data that currently lag in reconstruction performance \cite{cole2020unsupervised,Darestani2021AcceleratedMW,Gunel2021WS}, recent work has proposed designing data augmentation pipelines tailored to accelerated MRI reconstruction with appropriate image-based \cite{Fabian2021DataAF} or acquisition-based, physics-driven transformations \cite{desai2021noise2recon,Desai2022VORTEXPD}. Although helpful with data efficiency and robustness to certain distribution drifts, these approaches do not \emph{guarantee} that the final reconstruction model satisfies the desired symmetries, introduced through the data augmentation transformations, at train or inference time -- which may increase the existing concerns among clinicians around using data-driven techniques.\\\\
In this work, we propose modeling the proximal operators of unrolled neural networks with \textit{scale-equivariant} convolutional neural networks (CNNs) in order to improve data-efficiency and robustness to drifts in scale of the images that could be caused by the variability of patient anatomies or change in field-of-view across different MRI scanners. We note that our method effectively encodes a lack of prior knowledge of the scale of the structures in the images, in addition to the lack of position knowledge encoded by the \textit{translational equivariance} property of standard CNNs. Our approach \emph{ensures} more stable behavior for our model under scale and position changes at inference time, as scale and translation equivariance get explicitly encoded into the network. Here, we demonstrate the following:
\begin{itemize}
    \item Our approach outperforms state-of-the-art unrolled neural networks under the same memory constraints both with and without appropriate data augmentations on both in distribution and out-of-distribution scaled images with little impact on the train or inference time.
    \item Our method is empirically less sensitive to the step size initializations within the proximal updates in comparison to the state-of-the-art unrolled neural networks that are often tuned for each different dataset.
    \item We depict a correlation between the fidelity of enforcement of scale equivariance, quantified by equivariance error (Eq.~\ref{eq:equiv_error}), and reconstruction performance. We note that this reinforces the utility of incorporating the scale symmetry into unrolled neural networks to improve reconstruction performance in a data-efficient manner.
    \item We test our method on publicly available mridata 3D fast-spin-echo (FSE) multi-coil knee dataset \cite{ong2018mridata}. In order to promote reproducible research, we open-source our code, and will include experimental configurations along with trained models at \href{https://github.com/ad12/meddlr}{https://github.com/ad12/meddlr}.
\end{itemize}




%% file: content/2_related_work.tex
\textbf{Data Augmentation for Accelerated MRI Reconstruction.} Prior proposals designed transformations that leverage the natural symmetries of the accelerated MRI reconstruction problem in the form of data augmentation. Fabian et al. \cite{Fabian2021DataAF} proposed MRAugment, an image-based data augmentation pipeline. Desai et al. proposed a semi-supervised consistency framework for joint reconstruction and denoising \cite{desai2021noise2recon}, and later extended the denoising objective to a generalized data augmentation pipeline that enables composing a broader family of physics-driven acquisition-based augmentations and image-based augmentations \cite{Desai2022VORTEXPD}. \\\\
\textbf{Equivariant Networks.} Cohen et al. \cite{cohen2016group} showed that encoding symmetries directly into the neural network architectures using group equivariant CNNs lead to data-efficiency with guaranteed equivariance to encoded symmetries at both train and inference time. Following this work, there has been considerable amount of work in this direction across different domains of machine learning including exploring roto-translational symmetries \cite{Worrall2017HarmonicND,Cohen2018SphericalC} and scaling symmetries \cite{Worrall2019DeepSE,sosnovik2019scale,sosnovik2021disco}. Within medical applications, prior proposals primarily focused on roto-translational symmetries for both classification $\&$ segmentation tasks \cite{Bekkers2018RotoTranslationCC,Weiler20183DSC,Winkels20183DGF,Mller2021RotationEquivariantDL} and reconstruction tasks \cite{Chen2021EquivariantIL,Celledoni2021EquivariantNN}. To the best of our knowledge, there has not been any prior work that explored scale equivariance in the context of accelerated MRI reconstruction or for any other type of inverse problem.


%% file: content/3_background_and_preliminaries.tex
\subsection{Accelerated MRI Reconstruction}

We consider multi-coil MRI acquisition, which is a clinically-relevant setup where multiple receiver coils are used to acquire spatially-localized k-space measurements modulated by corresponding \textit{sensitivity maps}. In this setup, scan times are accelerated by decreasing the number of samples acquired in k-space, referred to as \textit{accelerated MRI reconstruction}. We represent the undersampling operation on acquired samples in k-space as a binary mask $\Omega$. Overall, the multi-coil accelerated MRI problem can be written as 
\begin{equation}
    y = \Omega\boldsymbol{F}\boldsymbol{S}x^* + \epsilon = Ax^* + \epsilon,
\end{equation}
where $y$ is k-space measurements, $\boldsymbol{F}$ is the matrix for discrete Fourier transform, $\boldsymbol{S}$ is the coil sensitivity maps, $x^*$ is the underlying ground truth image, and $\epsilon$ is the additive complex Gaussian noise. Coil sensitivity maps are estimated to perform reconstruction in practice as they are often unknown and vary per patient \cite{sense}. $A = \Omega\boldsymbol{F}\boldsymbol{S}$ is the known \textit{forward operator} during acquisition. Note that this problem is ill-posed in the Hadamard sense \cite{HadamardSurLP}, which makes recovering the underlying image $x^*$ impossible to recover uniquely without an assumption such as sparsity in a transformation domain as in compressed sensing \cite{cs-mri}.

\subsection{Unrolled Proximal Gradient Descent Networks}\label{sec:unrolled_pgd}
For MRI compressed sensing, the ill-posed reconstruction problem can be addressed with a regularized least-squares formulation of the form 
\begin{align}\label{eq:optimization_prob}
    \hat{x} = \arg\min_{x} \|Ax - y\|_2^2 + \mathcal{R}(x),
\end{align}
where \(\mathcal{R}\) is some regularizer (originally proposed an \(\ell_1\) penalty in the Wavelet domain \cite{cs-mri}). Problems of this form, where \(\mathcal{R}\) is not necessarily smooth, are often solved with iterative optimization methods such as Proximal Gradient Descent (PGD). At iteration \(k\), PGD operates as follows:
\begin{align}
    z^{(k)} &= x^{(k)} - \eta_k \nabla_{x}\|Ax^{(k)} - y\|_2^2 \label{eq:cons_step}\\
    x^{(k+1)} &= \mathrm{prox}_{\mathcal{R}}(z^{(k)}) \label{eq:prox_step}
\end{align}
with appropriately chosen step sizes \(\eta_k\), and proximal operator \(\mathrm{prox}_{\mathcal{R}}(\cdot)\). The first step thus takes a gradient step to enforce consistency of the iterate \(x^{(k)}\) with the measured signal in k-space \(y\) (data consistency), while the second step enforces the prior of regularizer \(\mathcal{R}\) on \(x^{(k)}\) (proximal step). It is known that following this procedure will provably solve Eq. \eqref{eq:optimization_prob} \cite{parikh2014proximal}. Choosing \(\mathcal{R}\) requires strong a priori assumptions that may not hold in practice. Thus, with the recent success of deep learning approaches, it has been proposed to replace the proximal step with a data-driven supervised learning approach: a learned neural network \cite{sandino2020compressed}. In particular, one unrolls a fixed, small number of iterations \(K\) of Eqs. \eqref{eq:cons_step} and \eqref{eq:prox_step}, and replaces each proximal step with a CNN, re-writing Eq. \eqref{eq:prox_step} with
\begin{equation}\label{eq:unrolled_prox}
    x^{(k+1)} = f_{\theta_k}(z^{(k)}). 
\end{equation}
The parameters \(\{\theta_k, \eta_k\}_{k=1}^K\) are then trained in an end-to-end fashion over a dataset of undersampled k-space and ground-truth image pairs \(\{y_i, x_i^*\}_{i=1}^n\) with some loss function \(\mathcal{L}(\hat{x}_i, x_i^*)\), such as the pixel-wise complex-\(\ell_1\) loss \(\|\hat{x}_i- x_i^*\|_1\). Such unrolled networks can outperform standard iterative compressed sensing methods, both in reconstruction quality and time \cite{sandino2020compressed}. One proposed explanation for the improved performance is that unrolled networks impose certain \emph{priors} that are not captured by \(\ell_1\)-Wavelet regularizers \cite{mardani2018neural,sahiner2020convex}. One such prior is that of \emph{translation equivariance}, which we describe in the subsequent section.

\subsection{Equivariance}
We say that a function \(f\) is \emph{equivariant} to a function \(g\) if
\begin{equation}\label{eq:equiv_defn}
    f(g(x)) = g(f(x))~\forall x
\end{equation}
Furthermore, we say \(f\) is equivariant to a \emph{family} of functions \(\mathcal{G}\) if Eq. \eqref{eq:equiv_defn} holds for all \(g \in \mathcal{G}\). Standard convolutional layers are equivariant to the discrete translation group --- if the input to a convolutional layer is translated in any direction, the corresponding output is the result of translating the original input's corresponding output. Thus, it is built into a CNN that translation does not affect the final output of the network (without pooling), thereby imposing an implicit prior on the nature of the functions that a CNN can learn. Note that in the context of unrolled proximal gradient descent networks, a translation equivariant proximal operator implies a translation invariant implicit regularizer \(\mathcal{R}\) \cite{Celledoni2021EquivariantNN}. One may also desire to impose other priors through equivariance, as have been proposed with rotation-equivariant CNNs \cite{cohen2016group} and scale-equivariant CNNs \cite{sosnovik2019scale}. 

%% file: content/4_methods.tex
\subsection{Learned Iterative Scale-Equivariant Reconstruction Networks}

We propose the use of \emph{scale-translation equivariant} CNNs for unrolled proximal gradient descent networks, referred to as \emph{unrolled neural networks}, for accelerated MRI reconstruction. \emph{In particular, we enforce the prior that either scaling or translating the undersampled input in the image domain should correspond to a scaled and translated output ground truth image.} This scale-translation equivariance provides additional built-in priors to the network, which creates resilience to the changes of scale in the images that might stem from the variability of patient anatomies or change in field-of-view across different MR scanners. We thus replace the proximal step in Eq. \eqref{eq:prox_step} of the unrolled proximal gradient descent network as described in Section \ref{sec:unrolled_pgd} with a scale-translation equivariant CNN as \(f_{\theta_k}\) in Eq. \eqref{eq:unrolled_prox}, leaving the data consistency steps unchanged. We refer to scale-translation equivariance as \emph{scale equivariance} throughout the text for brevity.

\subsection{Implementation}
We define the scale-translation group \(H = \{(s, t)\} := S \rtimes T\) as the semi-direct product of the scale group \(S\), which consists of scale transformations \(s\), and the translation group \(T\), which consists of translations \(t\) \cite{sosnovik2019scale}. A scale transformation \(L_{\hat{s}}\) of a function \(f\) defined on \(H\) is defined by scaling both the scale component and translation component of the arguments of \(f\), i.e.
\begin{equation}
    L_{\hat{s}} f(s, t) = f(s\hat{s}^{-1}, \hat{s}^{-1}t).
\end{equation}
For a convolutional kernel \(w\) to be scale-translation equivariant for a specific scale \(s\), we require, by Eq. \eqref{eq:equiv_defn}, 
\begin{equation}\label{eq:scale_conv_req}
    L_{s}[f] \star w = L_s[f \star w_{s^{-1}}],
\end{equation}
where \( w_{s^{-1}}\) is \(w\) scaled by \(s^{-1}\). To enforce scale equivariance, we follow \cite{sosnovik2019scale} with an implementation of scale-equivariant convolutions for discrete scales with steerable filters. A convolutional filter is scale-\emph{steerable} if an arbitrary scale transformation can be expressed as a linear combination of a fixed set of basis filters.  Each filter is expressed in terms of a \(B\)-dimensional \emph{kernel basis}, where each basis filter is pre-calculated and fixed. For our case, we desire that the convolution to be equivariant to scaling an image by a single factor \(a > 0\). We can then write our convolutional weights \(w\) as a linear combination of basis functions:
\begin{equation}
    w = \sum_{i=1}^{B} v_i \psi_i^{(a)},
\end{equation}
where \(v_i\) is learned, and \(\psi_i^{(a)}\) is a fixed basis filter which has been scaled by a factor of \(a^{-1}\). This method has proven to provide scale equivariance, since convolving an image scaled \(x\) by \(a\) with a filter \(w\) is equivalent to convolving \(x\) with a filter \(w\) scaled by \(a^{-1}\) and then subsequently downscaled, by Eq. \eqref{eq:scale_conv_req}. In our work, we choose \(\psi\) to be the set of 2D Hermite polynomials with a Gaussian envelope, which has shown empirically to work well in image classification and tracking tasks, though other bases, such as Fourier or radial, may also be used \cite{sosnovik2019scale,sosnovik2021disco}. We note that because these basis filters are based on continuous space and are then projected onto the pixel grid, the scale equivariance is not exact. This discrepancy between these two components for some function \(\Phi\), i.e.
\begin{equation}
    \Delta = \frac{\|L_s[\Phi(f)] - \Phi(L_s[f])\|_2^2}{\|L_s[\Phi(f)]\|_2^2}\label{eq:equiv_error}
\end{equation}
is defined as \emph{equivariance error} of \(\Phi\). This error suggests that in many cases, exact equivariance is not possible to enforce due to the discrete spatial nature of convolutional kernels. However, using the projection onto the pixel grid as described, one may find close approximations to exact equivariance. 

%% file: content/5_experiments.tex
\insertscalevsvanilla
\insertallequivariance
\insertmraugmentconfig
\insertreconexamples
\insertequiverror
\insertstepsize
\insertbenchmark

We use the publicly-available mridata 3D fast-spin-echo (FSE) multi-coil knee dataset (\href{http://mridata.org/}{http://mridata.org/}) \cite{ong2018mridata} of healthy patients to evaluate our approach. We decode 3D MRI scans into a hybrid k-space ($x \times k_y \times k_z$) using the 1D orthogonal inverse Fourier transform along the readout direction $x$ so that all methods reconstruct 2D $k_y \times k_z$ slices. We estimate the coil sensitivity maps using JSENSE \cite{ying2007joint} as implemented in SigPy \cite{sigpy} with a kernel width of 8 and a 20x20 center k-space auto-calibration region. We use 2D Poission Disc undersampling with a 16x acceleration rate, a compressed sensing motivated sampling pattern for 3D Cartesian imaging that has been routinely implemented in the clinic at Stanford Hospital \cite{Vasanawala2011PracticalPI}, for training and evaluation. We randomly partition the dataset into 4480 slices (14 volumes) for training, 640 slices (2 volumes) for validation, and 960 slices (3 volumes) for testing; and simulate a \textit{data-limited regime} with limiting the training set to 320 slices (1 volume). We use magnitude structural similarity (SSIM) \cite{wang2004image} and complex peak signal-to-noise ratio (cPSNR) in decibels (dB) to evaluate the quality of our reconstructions. We note that SSIM has shown to be a more clinically-preferred metric to cPSNR for quantifying perceptual quality of MRI reconstructions \cite{knoll2020advancing}. For all evaluated models, we train for 200 epochs, checkpoint every 10 epochs, and pick the best model based on validation SSIM; use the Adam optimizer \cite{kingma2014adam} with its default parameters ($\beta_1$=0.9, $\beta_2$=0.999, base learning rate of 1e-3); open-source (blinded) our implementations, experimental configurations, and trained models in PyTorch version 1.8.1 \cite{Paszke2019PyTorchAI}. \\\\
We follow Sandino et al. \cite{sandino2020compressed}'s implementation for the state-of-the-art unrolled neural networks that we refer to as \textit{Vanilla}. For the rotation-equivariant unrolled neural network baseline that we refer to as \textit{Rotation-Eq}, we follow Celledoni et al's \cite{Celledoni2021EquivariantNN} work in our own implementation where we enforce equivariance to $90^{\circ}$ rotations. For our scale-equivariant unrolled neural networks, referred to as \textit{Scale-Eq}, we set the scale parameter \(a=0.9\) and consider the order of the kernel basis a hyperparameter and set it to \(B=3\), unless specified otherwise. All unrolled neural networks use 5 unrolled blocks where each proximal block consists of 1 residual block that includes convolutions with 96 channels, learnable step sizes $\eta_k$, and pixel-wise complex-$\ell_1$ training objective. \\\\
We compare our method to baselines in the case where appropriate data augmentations are performed during training in order to demonstrate the practical utility of building equivariant networks \textit{in addition to} the data augmentation based approaches. Specifically, we refer to Vanilla as Vanilla+ and Scale-Eq as Scale-Eq+ when scale data augmentations with a scaling factor of \(a=0.9\) are introduced with simple exponential augmentation probability scheduling \cite{Fabian2021DataAF}. In Table~\ref{table:maintable}, we consider the cases where learnable step sizes (described in Eq.~\ref{eq:cons_step}) are zero-initialized or tuned as a hyperparameter, and evaluate on both regular test slices and unseen scaled test slices with a scaling factor of \(a=0.9\) in the data-limited regime. Taking both target ground truth images and pre-computed coil sensitivity maps into account, the scaled setting simulates the scaling changes at inference time which might stem from the variability of patient anatomies or change of field-of-view across different MRI scanners in practice. We show that Scale-Eq+ outperforms Vanilla+ across all settings, overall considerably improving the data-efficiency of the network. We note that we keep the number of trainable parameters almost the same between Scale-Eq+ and Vanilla+ for fair performance comparison and demonstrate that Scale-Eq+ does not significantly increase the training or inference time over Vanilla+ in Table 5 to ensure practical utility. We include reconstruction examples along with error maps for Vanilla+ and Scale-Eq+ in Figure 1. We empirically demonstrate in Table 4 that Scale-Eq is less sensitive than Vanilla to the learnable step size initializations that is often highly specific to the dataset and hence can require careful tuning. \\\\
In Table~\ref{table:allequivariance}, we compare Vanilla, Rotation-Eq, and Scale-Eq using the state-of-the-art data augmentation pipeline MRAugment \cite{Fabian2021DataAF}, which includes image-based pixel-preserving augmentations such as translation, arbitrary and 90 degree multiple rotations, translation, as well as isotropic and anisotropic scaling, in its default configuration specified in Table 3. We demonstrate that Scale-Eq-MRAugment outperforms Vanilla-MRAugment while Rotation-Eq-MRAugment outperforms Vanilla-MRAugment in PSNR and does comparably in SSIM. We clarify that we do not aim to argue that scale symmetry is more useful to encode than rotation symmetry here, in fact, we demonstrate that encoding scale symmetry is as helpful as encoding rotation symmetry, if not more. Finally, in Figure 2, we depict a correlation between the fidelity of enforcement of scale equivariance, quantified by equivariance error (Eq.~\ref{eq:equiv_error}), and reconstruction performance in terms of SSIM. This further reinforces the utility of incorporating scale equivariance into unrolled networks to improve reconstruction performance in a data-efficient manner.


%% file: content/6_conclusion.tex
We have proposed learned iterative scale-equivariant unrolled neural networks for data-efficient accelerated MRI reconstruction, demonstrating its utility in improving reconstruction performance in both in-distribution and out-of-distribution settings, even under cases of data augmentation during training. We thus demonstrate that encoding the lack of prior knowledge of the scale of the images can provide a more robust reconstruction network. Many other directions related to this work can be explored, such as applications to three-dimensional image recovery, or to other medical image tasks such as segmentation. One may also further explore the impact of particular transformation groups to which one may be equivariant, to encode multiple priors into various image processing tasks. 

%% file: content/7_acknowledgements.tex
Beliz Gunel, Arda Sahiner, Shreyas Vasanawala, and John Pauly were supported by NIH R01EB009690 and NIH U01-EB029427-01. Mert Pilanci was partially supported by the National Science Foundation under grants IIS-1838179, ECCS- 2037304, DMS-2134248, and the Army Research Office. Arjun Desai and Akshay Chaudhari were supported by grants R01 AR077604, R01 EB002524, K24 AR062068, and P41 EB015891 from the NIH; the Precision Health and Integrated Diagnostics Seed Grant from Stanford University; National Science Foundation (DGE 1656518, CCF1763315, CCF1563078); DOD -- National Science and Engineering Graduate Fellowship (ARO); Stanford Artificial Intelligence in Medicine and Imaging GCP grant; Stanford Human-Centered Artificial Intelligence GCP grant; Microsoft Azure through Stanford Data Science’s Cloud Resources Program; GE Healthcare and Philips.

%% file: miccai2022.bbl
\begin{thebibliography}{10}
\providecommand{\url}[1]{\texttt{#1}}
\providecommand{\urlprefix}{URL }
\providecommand{\doi}[1]{https://doi.org/#1}

\bibitem{Aggarwal_Mani_Jacob_2019}
Aggarwal, H.K., Mani, M.P., Jacob, M.: Modl: Model-based deep learning
  architecture for inverse problems. IEEE transactions on medical imaging
  \textbf{38}(2),  394--405 (2018)

\bibitem{Bekkers2018RotoTranslationCC}
Bekkers, E.J., Lafarge, M.W., Veta, M., Eppenhof, K.A.J., Pluim, J.P.W., Duits,
  R.: Roto-translation covariant convolutional networks for medical image
  analysis. ArXiv  \textbf{abs/1804.03393} (2018)

\bibitem{Celledoni2021EquivariantNN}
Celledoni, E., Ehrhardt, M.J., Etmann, C., Owren, B., Schonlieb, C.B., Sherry,
  F.: Equivariant neural networks for inverse problems. Inverse Problems
  \textbf{37} (2021)

\bibitem{Chaudhari2021}
Chaudhari, A.S., Sandino, C.M., Cole, E.K., Larson, D.B., Gold, G.E.,
  Vasanawala, S.S., Lungren, M.P., Hargreaves, B.A., Langlotz, C.P.:
  Prospective deployment of deep learning in mri: A framework for important
  considerations, challenges, and recommendations for best practices. Journal
  of Magnetic Resonance Imaging  (2020)

\bibitem{Chen2021EquivariantIL}
Chen, D., Tachella, J., Davies, M.E.: Equivariant imaging: Learning beyond the
  range space. ArXiv  \textbf{abs/2103.14756} (2021)

\bibitem{Cohen2018SphericalC}
Cohen, T., Geiger, M., K{\"o}hler, J., Welling, M.: Spherical cnns. ArXiv
  \textbf{abs/1801.10130} (2018)

\bibitem{cohen2016group}
Cohen, T., Welling, M.: Group equivariant convolutional networks. In:
  International conference on machine learning. pp. 2990--2999. PMLR (2016)

\bibitem{cole2020unsupervised}
Cole, E.K., Pauly, J.M., Vasanawala, S.S., Ong, F.: Unsupervised mri
  reconstruction with generative adversarial networks. arXiv preprint
  arXiv:2008.13065  (2020)

\bibitem{Darestani2021AcceleratedMW}
Darestani, M.Z., Heckel, R.: Accelerated mri with un-trained neural networks.
  IEEE Transactions on Computational Imaging  \textbf{7},  724--733 (2021)

\bibitem{Desai2022VORTEXPD}
Desai, A.D., Gunel, B., Ozturkler, B.M., Beg, H., Vasanawala, S., Hargreaves,
  B.A., R{\'e}, C., Pauly, J.M., Chaudhari, A.S.: Vortex: Physics-driven data
  augmentations for consistency training for robust accelerated mri
  reconstruction. In: MIDL (2022)

\bibitem{desai2021noise2recon}
Desai, A.D., Ozturkler, B.M., Sandino, C.M., Vasanawala, S., Hargreaves, B.A.,
  Re, C.M., Pauly, J.M., Chaudhari, A.S.: Noise2recon: A semi-supervised
  framework for joint mri reconstruction and denoising. arXiv preprint
  arXiv:2110.00075  (2021)

\bibitem{Fabian2021DataAF}
Fabian, Z., Heckel, R., Soltanolkotabi, M.: Data augmentation for deep learning
  based accelerated mri reconstruction with limited data. In: International
  Conference on Machine Learning. pp. 3057--3067. PMLR (2021)

\bibitem{Gunel2021WS}
Gunel, B., Mardani, M., Chaudhari, A., Vasanawala, S., Pauly, J.: Weakly
  supervised mr image reconstruction using untrained neural networks. In:
  Proceedings of International Society of Magnetic Resonance in Medicine
  (ISMRM) (2021)

\bibitem{HadamardSurLP}
Hadamard, J.: Sur les probl{\`e}mes aux d{\'e}riv{\'e}es partielles et leur
  signification physique. Princeton university bulletin pp. 49--52 (1902)

\bibitem{kingma2014adam}
Kingma, D.P., Ba, J.: Adam: A method for stochastic optimization. arXiv
  preprint arXiv:1412.6980  (2014)

\bibitem{knoll2020advancing}
Knoll, F., Murrell, T., Sriram, A., Yakubova, N., Zbontar, J., Rabbat, M.,
  Defazio, A., Muckley, M.J., Sodickson, D.K., Zitnick, C.L., et~al.: Advancing
  machine learning for mr image reconstruction with an open competition:
  Overview of the 2019 fastmri challenge. Magnetic resonance in medicine
  \textbf{84}(6),  3054--3070 (2020)

\bibitem{lustig2007sparse}
Lustig, M., Donoho, D., Pauly, J.M.: Sparse mri: The application of compressed
  sensing for rapid mr imaging. Magnetic Resonance in Medicine: An Official
  Journal of the International Society for Magnetic Resonance in Medicine
  \textbf{58}(6),  1182--1195 (2007)

\bibitem{cs-mri}
Lustig, M., Donoho, D.L., Santos, J.M., Pauly, J.M.: Compressed sensing mri.
  IEEE signal processing magazine  \textbf{25}(2),  72--82 (2008).
  \doi{10.1109/MSP.2007.914728}

\bibitem{mardani2018neural}
Mardani, M., Sun, Q., Donoho, D., Papyan, V., Monajemi, H., Vasanawala, S.,
  Pauly, J.: Neural proximal gradient descent for compressive imaging. Advances
  in Neural Information Processing Systems  \textbf{31} (2018)

\bibitem{Mller2021RotationEquivariantDL}
M{\"u}ller, P., Golkov, V., Tomassini, V., Cremers, D.: Rotation-equivariant
  deep learning for diffusion mri. ArXiv  \textbf{abs/2102.06942} (2021)

\bibitem{ong2018mridata}
Ong, F., Amin, S., Vasanawala, S., Lustig, M.: Mridata.org: An open archive for
  sharing mri raw data. In: Proc. Intl. Soc. Mag. Reson. Med. vol.~26, p.~1
  (2018)

\bibitem{sigpy}
Ong, F., Lustig, M.: Sigpy: a python package for high performance iterative
  reconstruction. In: Proceedings of the ISMRM 27th Annual Meeting, Montreal,
  Quebec, Canada. vol.~4819 (2019)

\bibitem{parikh2014proximal}
Parikh, N., Boyd, S.: Proximal algorithms. Foundations and Trends in
  optimization  \textbf{1}(3),  127--239 (2014)

\bibitem{Paszke2019PyTorchAI}
Paszke, A., Gross, S., Massa, F., Lerer, A., Bradbury, J., Chanan, G., Killeen,
  T., Lin, Z., Gimelshein, N., Antiga, L., Desmaison, A., K{\"o}pf, A., Yang,
  E., DeVito, Z., Raison, M., Tejani, A., Chilamkurthy, S., Steiner, B., Fang,
  L., Bai, J., Chintala, S.: Pytorch: An imperative style, high-performance
  deep learning library. In: NeurIPS (2019)

\bibitem{sense}
Pruessmann, K.P., Weiger, M., Scheidegger, M.B., Boesiger, P.: Sense:
  sensitivity encoding for fast mri. Magnetic Resonance in Medicine: An
  Official Journal of the International Society for Magnetic Resonance in
  Medicine  \textbf{42}(5),  952--962 (1999)

\bibitem{sahiner2020convex}
Sahiner, A., Mardani, M., Ozturkler, B., Pilanci, M., Pauly, J.: Convex
  regularization behind neural reconstruction. arXiv preprint arXiv:2012.05169
  (2020)

\bibitem{sandino2020compressed}
Sandino, C.M., Cheng, J.Y., Chen, F., Mardani, M., Pauly, J.M., Vasanawala,
  S.S.: Compressed sensing: From research to clinical practice with deep neural
  networks: Shortening scan times for magnetic resonance imaging. IEEE signal
  processing magazine  \textbf{37}(1),  117--127 (2020)

\bibitem{sosnovik2021disco}
Sosnovik, I., Moskalev, A., Smeulders, A.: Disco: accurate discrete scale
  convolutions. arXiv preprint arXiv:2106.02733  (2021)

\bibitem{sosnovik2019scale}
Sosnovik, I., Szmaja, M., Smeulders, A.: Scale-equivariant steerable networks.
  arXiv preprint arXiv:1910.11093  (2019)

\bibitem{Sriram2020EndtoEndVN}
Sriram, A., Zbontar, J., Murrell, T., Defazio, A., Zitnick, C.L., Yakubova, N.,
  Knoll, F., Johnson, P.: End-to-end variational networks for accelerated mri
  reconstruction. In: International Conference on Medical Image Computing and
  Computer-Assisted Intervention. pp. 64--73. Springer (2020)

\bibitem{Vasanawala2011PracticalPI}
Vasanawala, S.S., Murphy, M.J., Alley, M.T., Lai, P., Keutzer, K., Pauly, J.M.,
  Lustig, M.: Practical parallel imaging compressed sensing mri: Summary of two
  years of experience in accelerating body mri of pediatric patients. 2011 IEEE
  International Symposium on Biomedical Imaging: From Nano to Macro pp.
  1039--1043 (2011)

\bibitem{wang2004image}
Wang, Z., Bovik, A.C., Sheikh, H.R., Simoncelli, E.P.: Image quality
  assessment: from error visibility to structural similarity. IEEE transactions
  on image processing  \textbf{13}(4),  600--612 (2004)

\bibitem{Weiler20183DSC}
Weiler, M., Geiger, M., Welling, M., Boomsma, W., Cohen, T.: 3d steerable cnns:
  Learning rotationally equivariant features in volumetric data. In: NeurIPS
  (2018)

\bibitem{Winkels20183DGF}
Winkels, M., Cohen, T.: 3d g-cnns for pulmonary nodule detection. ArXiv
  \textbf{abs/1804.04656} (2018)

\bibitem{Worrall2017HarmonicND}
Worrall, D.E., Garbin, S.J., Turmukhambetov, D., Brostow, G.J.: Harmonic
  networks: Deep translation and rotation equivariance. 2017 IEEE Conference on
  Computer Vision and Pattern Recognition (CVPR) pp. 7168--7177 (2017)

\bibitem{Worrall2019DeepSE}
Worrall, D.E., Welling, M.: Deep scale-spaces: Equivariance over scale. ArXiv
  \textbf{abs/1905.11697} (2019)

\bibitem{ying2007joint}
Ying, L., Sheng, J.: Joint image reconstruction and sensitivity estimation in
  sense (jsense). Magnetic Resonance in Medicine: An Official Journal of the
  International Society for Magnetic Resonance in Medicine  \textbf{57}(6),
  1196--1202 (2007)

\end{thebibliography}
